\documentclass[doublecol]{epl2} 

\usepackage{color}
\usepackage{soul}                    
\usepackage{ulem}                    
\usepackage{graphics,graphicx,dcolumn,bm,fleqn,epic,eepic,float}
\usepackage{amssymb,amsmath,multirow,rotate,color,float,times}

\definecolor{red}{rgb}{1,0,0}
\definecolor{green}{rgb}{0,1,0}
\definecolor{blue}{rgb}{0,0,1}

\title{A thermostatistical approach to scale-free networks}

\author{Jo\~ao P. da Cruz \inst{1,2} Nuno A. M. Ara\'ujo \inst{2}
        Frank Raischel \inst{3} Pedro G. Lind \inst{4}}
\institute{
\inst{1} Closer Consultoria Lda, Avenida Engenheiro Duarte Pacheco,
         Torre 2, 
         1070-102 Lisboa, Portugal\\
\inst{2} Center for Theoretical and Computational Physics, 
         University of Lisbon, 
         1649-003 Lisbon, Portugal\\
\inst{3} Instituto Dom Luiz, CGUL, 1749-016 University of Lisbon, 
         Lisbon, Portugal\\
\inst{4} ForWind and Institute of Physics, 
         University of Oldenburg, 
         DE-26111 Oldenburg, Germany
}


\abstract{
We describe an ensemble of growing scale-free networks in an equilibrium 
framework, providing insight into why the exponent of empirical 
scale-free networks in nature is typically robust.  
In an analogy to thermostatistics, to describe the canonical and 
microcanonical ensembles, we introduce a functional, whose maximum 
corresponds to a scale-free configuration.
We then identify the equivalents to energy, Zeroth-law, entropy and 
heat capacity for scale-free networks.
Discussing the merging of scale-free networks, we also establish an 
exact relation to predict their final ``equilibrium'' degree exponent. 
All analytic results are complemented with Monte Carlo simulations.
Our approach illustrates the possibility to apply the tools of equilibrium 
statistical physics to study the properties of growing networks, and
it also supports the recent arguments on the complementarity between 
equilibrium and nonequilibrium systems.
} 

\begin{document}
\maketitle

\section{Introduction}

The theory of thermal equilibrium is one of the most important milestones 
in the development of statistical and theoretical physics.
Most mesoscopic processes are, however, far from equilibrium,
and describe the flow of energy through a system. Often the cascaded transfer 
and eventual dissipation of this energy leads to the appearance of scale-free
structures, characterized by power laws.
Describing 
nonequilibrium systems in a equilibrium framework would give access to 
the vast array of tools developed for the equilibrium case.
Such paralelism is not straightforward. Nevertheless,
important equivalences between the two descriptions
have recently been established~\cite{chetrite2013,marco2013}.

With the aim of strengthen further the bridge between equilibrium and
non-equilibrium systems, we explore such apparent
equivalences in the context of complex networks.
Networks are general means to describe the topology of interactions,
where nodes are the agents and the links represent the interactions.
Scale-free networks (SFNs) are characterized by a power-law degree
distribution and have been found in several socio-technical systems,
such as airline networks, Internet, phone calls patterns, and reservoir
networks~\cite{guimera2005,barabasi,marta2008,hansPNAS}. Insofar, SFNs
have been described as products of growth processes as, e.g.,
preferential attachment~\cite{Boccaletti2006,Castellano2009}, and
therefore classified as nonequilibrium
networks~\cite{DorogovtsevRMP2008}. 

In this Letter, we present a mapping
of a specific ensemble of SFNs into an equilibrium ensemble 
by properly rescaling the growing values of the connection strength,
and we apply the tools of thermostatistics to measure several 
network properties. We derive the equivalents to micro- and macrostates 
and show how properties such as energy and entropy can be defined.
While the number of nodes is kept constant, the number of connections
grows in time.

For the specific case that the weights correspond to the degrees $k$ of
the nodes, we present a series of Monte-Carlo simulations,
both in the canonical and the microcanonical
formalism, illustrating that the scale-free networks follow  the
equivalent of a Zeroth law, i.~e.~{\it (i)} a network brought in contact
with a reservoir relaxes to an ``equilibrium'' state with the same
degree distribution as the reservoir, {\it (ii)} two ``equilibrated''
networks with different degree distributions brought in contact with
each other develop a new ``equilibrium'' state with a new final degree
distribution. We then show, {\it (iii)}, that the final degree
distribution can be calculated exactly from the equivalent of the heat
capacity, which depends on the second moment of the energy. 
\begin{figure}[t]
\centerline{\includegraphics[width=0.89\columnwidth]{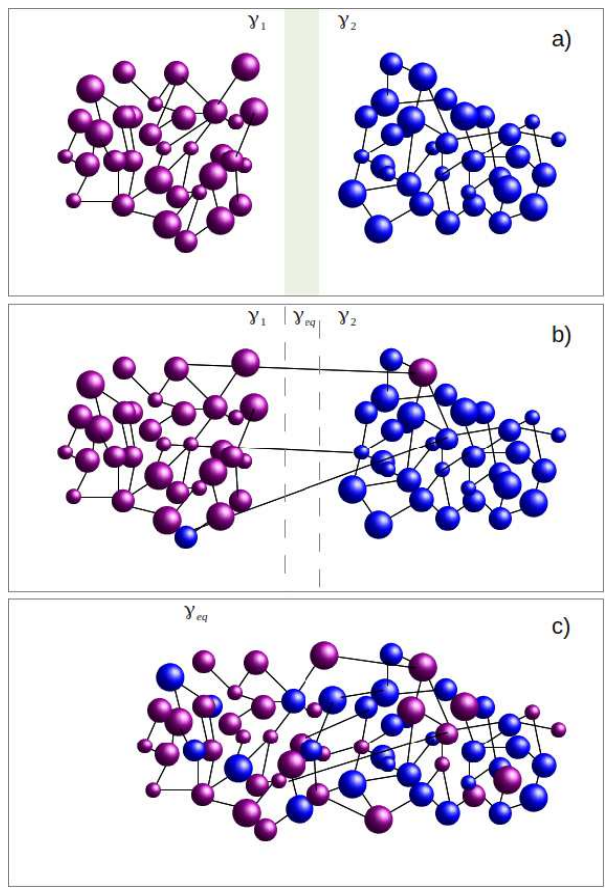}}
\caption{\protect (Color online) When two scale-free networks of degree
exponents $\gamma_1$ and $\gamma_2$ are put in contact, they evolve
towards an ``equilibrium'' scale-free state of exponent $\gamma_{eq}$.
\label{fig01}} \end{figure}

We start by defining a proper energy function that characterizes the
microstates of a scale-free network. From this definition we then
introduce the canonical and the microcanonical ensembles. From these
ensembles we can then derive the explict expression for entropy and heat
capacity as well as principles such as the Zeroth-law for scale-free
networks. Discussion and conclusions end this paper.

\section{The energy function for scale-free networks}

The network here investigated consider an ensemble of $N$ nodes, where
each node $i$ occupies one of $M$ possible microstates. Each microstate $j$
is characterized by a scalar $\hat w_j$, which we call connection strength
or weight. 
This weight can be, for example, the degree of the node for undirected
graphs, the in- or out-degree for directed ones, or the sum over the
weight of the node links for a weighted graph. 
In every case, it is
the weight that measures the local state of each node in the system, and
therefore the energy function shall be a function of the weight alone.

Since weights grow in time, the state space is not constant, and thus
one cannot denominate a statistical ensemble. To overcome that shortcoming
we define a rescaled weight $w_j$ as,
\begin{equation}\label{eq::normalized.omega}
w_j=\frac{\hat w_j-\hat w_{\textrm{min}}}{\hat w_{\textrm{max}}-\hat
w_{\textrm{min}}} \ \ ,
\end{equation}
where $\hat w_{\textrm{min}}$ and $\hat w_{\textrm{max}}$ are the
minimum and maximum weights, such that $w_j$ is in the range $[0,1]$,
for all states $j=1,\dots,M$. 
With this rescaling, we then have a time-independent probability space, 
over which we can derive a microcanonical evolution equation. 

Assuming that the weights are an equivalent of an energy flow into 
the system, the next question is: 
Which functional dependence on the weight should the energy function have?
Here we argue that the proper choice for the energy function 
of the microstate $j$ is
\begin{equation}\label{eq::energy}
\epsilon_j=\log{w_j} .
\end{equation}

Why is the
logarithmic function of the weight the proper one for defining an energy
interchange among nodes in a scale-free network?
The answer might lie in the typical self-similarity of scale-free
networks\cite{barabasi}.  In time-evolving processes one usually
looks for periodicities, i.e.~invariants under a given time shift.  For
an ordered set of different scales the corresponding feature of
periodicity is self-similarity\cite{schroederBook}: a perfectly
self-similar process variable $w$ is invariant when shifting from one
scale  to another, $\lambda$, i.e.~the probability
distribution function remains invariant $p(\lambda^n \omega ) = \lambda
p(\omega)$, as do its moments.

Through the rescaling 
of the weights in scale-free networks one is able to obtain a stationary 
probability space, in which the evolution of canonical and microcanonical 
ensembles can be formulated.
For that, it is necessary to identify the logarithmic weight as the
energy function of each microstate.
In other words, while SFNs evolve with no conservation of their
average degree, the total logarithmic weight is conserved,
which should be taken therefore as the total energy.

\section{Canonical scale-free networks}

To derive the canonical ensemble of scale-free networks we first
define each macrostate of the system.
One macrostate $W$ is completely defined by the distribution of the $N$
nodes among the microstrates: $W\equiv W(n_1,\dots,n_M)$, where $n_j$
is the occupation number of state $j$. For each macrostate there are
$\Omega=\frac{N!}{\prod_{j}n_j!}$ equivalent configurations or 
microstates. Using Boltzmann's entropy, the entropy $S$ of a 
macrostate is $S=\log{\Omega}$. 

Having two of such systems (networks), as sketched in
Fig.~\ref{fig01}, we now consider that they
evolve in time, i.e.~connections are created and destroyed in time
according to some criteria, reflecting an interchange of energy among
the nodes which, consequently hop between microstates. We next
introduce the canonical ensemble, which describes the distribution among
microstates for one single network, taking the other
one in Fig.~\ref{fig01} as a heat reservoir. After that we discuss
the microcanonical ensemble where both networks
are of the same size.

\begin{figure}[t]
\centerline{\includegraphics[width=\columnwidth]{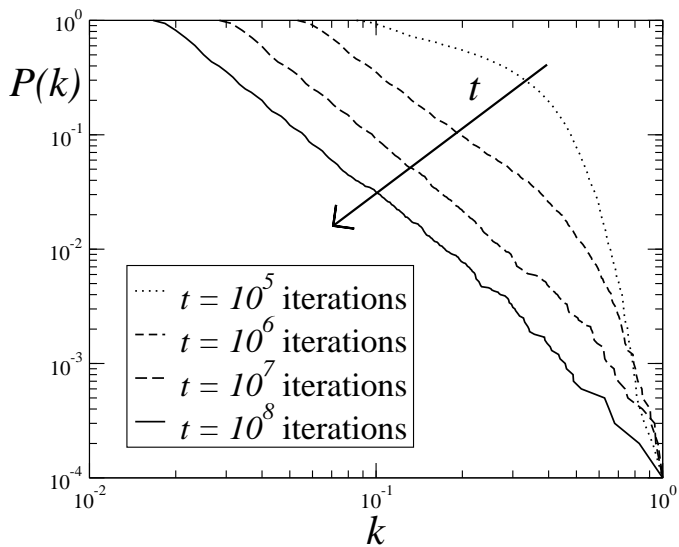}}
\caption{\protect The scale-free canonical ensemble: starting with a 
uniform distribution of degree among $N=10^5$ nodes in contact with a
reservoir ($N=10^7$ nodes) having a distribution $P(k)\sim k^{-\gamma}$ with
$\gamma=3$ , the degree distribution of the system evolves towards 
the same power-law.
\label{fig02}}
\end{figure}

Figure \ref{fig02} illustrates the evolution of a canonical scale-free
ensemble. We started with a regular random graph with $N=10^5$ nodes,
each one having $k=5$ neighbors, and put it in contact with a large 
scale-free network (reservoir) with $N=10^7$ nodes and exponent
$\gamma=3$. The latter was obtained using the described
algorithm after a sufficiently large number of interactions.
As one clearly sees throughout the four stages plotted in Fig.~\ref{fig02}, 
the degree distribution evolves towards a power-law distribution with
the same exponent.
While in this implementation the number $L$ of links is kept constant,
similar results are obtained for a network with $L$ growing in time;
in that case, instead of degree, one considers the normalized degree similar
to the weight in Eq.~(\ref{eq::normalized.omega}).

In a more general canonical framework, where one has a weight defined by
Eq.~(\ref{eq::normalized.omega}),
the most probable macrostate is the one that maximizes the
entropy. This maximization has two constraints: conservation of the
average total energy, 
which with the definition \eqref{eq::energy} reads $E=\sum_j n_j\log{w_j}$, 
and of the number of nodes, $N=\sum_j n_j$.  

Together with the definition of entropy above, 
the most probable macrostate is an
extremum of the functional,
\begin{equation}\label{eq::functional}
 {\cal{F}} = \log{\Omega} - \gamma \sum_{j=1}^M n_j\log{w_j} +
\log{\alpha} \sum_{j=1}^M n_j,
\end{equation}
where $\gamma$ and $\log{\alpha}$ are the Lagrange multipliers for the
constraints in the energy and number of nodes, respectively. For
thermodynamic ensembles, $\gamma$ is the inverse of the reservoir
``temperature'' and $\log{\alpha}$ the ratio between the 
``chemical potential'' and the temperature. 
Interestingly, from this approach one concludes that $\gamma$ is a 
property of the reservoir, while $\log{\alpha}$ is characteristic 
of the system.

Using  Stirling's approximation one obtains that the extrema of
Eq.~(\ref{eq::functional}) yield $n_j=\alpha w_j^{-\gamma}$. If we
define $\hat{w}_i=k_i$, where $k_i$ is the node degree, the degree
distribution is a power law of degree exponent $\gamma$. In other 
words, the reservoir sets the degree exponent. 
The weight distribution is consistent with the expected Boltzmann 
distribution for canonical ensembles. Note that the node energy 
is $\epsilon_i=\log{w_i}$, which is thereby exponentially 
distributed. 

In fact, as claimed by other authors, if properly interpreted, the
Boltzmann distribution yield ``disguised'' power laws
\cite{Solomon1996,richmond2001}.  In our case, the power law reflects
the degree distribution $P(k)\sim k^{-\gamma}$ of the connections among
the nodes of the system in contact with a ``heat'' reservoir having a
constant ``temperature'', i.e.~a very large 
network with a given exponent for its power-law degree distribution.
Notice that, the subnetwork converge to a SFN by randomly interchanging
connections with the reservoir: no preferential attachment is here 
required. One therefore obtains a SF topology from an arbitrary
initial weight distribution and a succession of random exchanges of 
connections with a SF reservoir. See Fig.~\ref{fig02}.

To generate such ``equilibrium'' networks with degree exponent
$\gamma$, the canonical formalism was numerically implemented as follows.
One starts with $N$ nodes, interconnected through $L$ links,
yielding a certain degree distribution, $\{k_i\}$.  At each iteration,
one randomly selects two nodes, $l$ and $m$, and calculates the change
$\Delta E$ in the total energy $E$ if one link connected to $l$ is 
rewired to be connect to $m$,
\begin{equation}\label{eq::energy.change}
\Delta E=\Delta(\log{k_l})+\Delta(\log{k_m}) \ \ .
\end{equation}
This rewiring is executed with probability
$p=\min\left\{1,\exp\left(-\gamma\Delta E\right)\right\}$. After some
iterations, the network converges to the desired scale-free degree
distribution. Since
$\epsilon_i=\log{k_i}$, the energy is only conserved for movements where
$k^{\mathrm{initial}}_l=k^{\mathrm{initial}}_m+1$, where $k^{\mathrm{initial}}_l$ 
and $k^{\mathrm{initial}}_m$ are the degree of nodes $l$ and $m$ before 
the movement. This algorithm is an instance  of the Metropolis
algorithm, thus respects detailed balance\cite{metropolis}.
\begin{figure}[t]
\centerline{\includegraphics[width=0.9 \columnwidth]{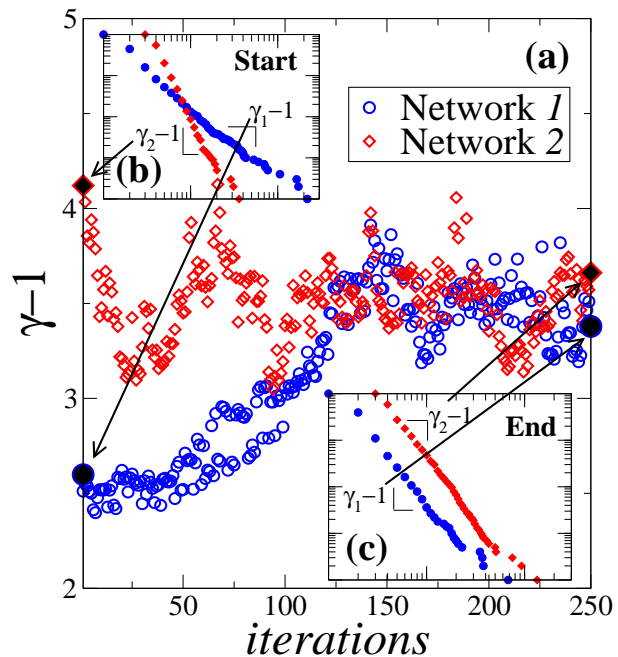}}
\caption{(Color online) (a) Evolution of the degree exponent of two
scale-free networks in contact with each other. 
Initially, the two networks have
different degree exponents (inset (b)) but
converge towards a degree distribution with the same degree exponent
$\gamma_\mathrm{eq}$ (inset (c)). All represented values refer to 
cumulative distributions.
\label{fig03}}
\end{figure}

\section{Microcanonical scale-free networks}

We consider now the classical problem of thermal equilibration. What is
the final degree distribution when two scale-free networks are put
in
contact? To address this question, as schematically illustrated in
Fig.~\ref{fig01}, we consider two scale-free networks of degree
exponent $\gamma_1$ and $\gamma_2$.  Since the combined system is
isolated, the total energy $E_{1+2}=E_1+E_2$ is expected to be
conserved, where $E_1$ and $E_2$ is the energy of networks $1$ and $2$,
respectively.  Figure~\ref{fig03}a shows the evolution of the degree
exponent of the nodes in each network, using the microcanonical
algorithm proposed  below.  The two networks merge into one with a
power-law degree distribution of exponent $\gamma_\mathrm{eq}$. 

This situation is described through a microcanonical ensemble, which
represents an isolated network at a given energy. To numerically explore
the configuration space we propose a generalization of the Creutz
algorithm, which respects both detailed balance and
ergodicity\cite{creutz1983}. Let us consider again the case where the
weight corresponds to the degree.  Algorithmically, instead of
interacting with a heat reservoir like in the canonical case, the
network exchanges energy with one energy reservoir, named demon, with
energy $E_\mathrm{demon}$. One starts with one network at the desired
energy and $E_\mathrm{demon}=0$. As in the canonical simulation, at each
iteration two nodes $l$ and $m$ are randomly selected and the energy
change of the rewiring movement is obtained from
Eq.~(\ref{eq::energy.change}). If $\Delta E\leq0$, the movement is
accepted and the energy difference is transferred to the demon.
Otherwise, the movement is only accepted if $E_\mathrm{demon}>\Delta E$,
and, when it happens, the energy difference is subtracted from 
$E_\mathrm{demon}$. To obtain a network with a pre-given energy $E$, one 
can also start with the subnetwork having the desired number of links and 
nodes and energy $E_{1+2}<E$ and impose $E_\textrm{demon}=E-E_0$. 
After a few iterations the network energy converges towards $E$.

To illustrate the microcanonical ensemble of scale-free networks we
implemented the algorithm above for two subnetworks,
each one composed of $N=10^5$ nodes and a power-law degree
distribution, but having different exponents, namely $\gamma_1\approx 7/2$ and
$\gamma_2\approx 5$.  
Figure~\ref{fig03}b shows the initial degree distribution of each
subnetwork. 
As shown in Figs.~\ref{fig03}b and \ref{fig03}c, the degree distribution 
of the set of nodes initially belonging to subnetworks 1 and 2 converges 
towards the same degree exponent, $\gamma_{eq}\approx 9/2$.

\section{Thermostatistics of scale-free networks}

With the thermostatistic approach proposed here,
one can derive five results for ensembles of scale-free networks. 
First,
the equivalent to the Zeroth-law of thermodynamics can be
expressed as: if networks $A$ and $B$ are in equilibrium with network
$C$ then they are also in equilibrium with each other. Here, two
networks in equilibrium have the same degree exponent $\gamma$.

Second, one can derive a closed expression for the entropy of scale-free
networks. As discussed before, each macrostate has $\Omega$ possible
equivalent configurations. Substituting the power-law solution in
Eq.~(\ref{eq::functional}) in the definition of $\Omega$, yields for 
the entropy 
\begin{equation}\label{eq::entropy}
  S=\log{\Omega}=S_0+\gamma \alpha \sum _{i=1} ^M { w_i
^{-\gamma}\log{w_i}} \ \ ,
\end{equation}
where $S_0 = N\log{\left ( \sum_{i=1}^N w_i^{-\gamma} \right )}$ is the
minimum entropy and the second term is equivalent to the Shannon
entropy. This expression for the entropy complements previous
works\cite{park2004}.

Third, in the continuum limit, associated with a large number of
microstates $M$, the sum in Eq.~(\ref{eq::entropy}) can be approximated
by an integral, yielding an expression for $S$, solely depending on
$\gamma$ and $N$. The differential of the entropy is then,
\begin{equation}\label{dSdgamma}
dS = \frac{\partial S}{\partial\gamma} d\gamma + \frac{\partial S}{\partial N} 
dN \ \ .
\end{equation}
Equation (\ref{dSdgamma}) tell us that the network entropy can either 
change due to removal or addition of nodes ($dN\neq 0$), similar to the 
procedures introduced by Barabasi, Newman and others\cite{barabasi} 
or due to rewiring\cite{Schneider2011}, ``energy'' exchange ($d\gamma\neq0$), 
as reported here.
\begin{figure}[t]
\centerline{\includegraphics[width=0.99\linewidth]{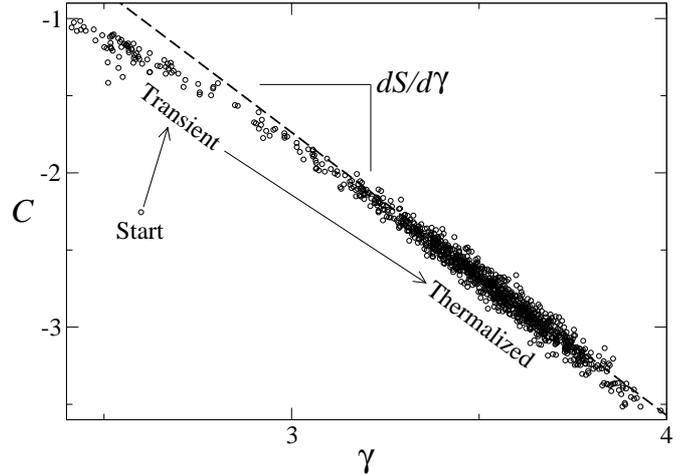}}
\vspace*{8pt}
\caption{\protect
         The heat capacity $c_N$ versus exponent $\gamma$ for the microcanonical
         ensemble of two scale-free networks. The negative-valued slope of the
         linear dependence between $c$ and $\gamma$ gives the constant value
         $\frac{dS}{d\gamma}$. See Fig.~\ref{fig05}.} 
\label{fig04}
\end{figure}

Fourth, we can define a heat capacity for non-growing scale-free networks, 
which can be obtained from the entropy as,
\begin{equation}\label{eq::heat.capacity}
c_N = \gamma \left (\frac{dS}{d\gamma}\right )_{N} = \gamma\langle 
\log{w}\rangle - \gamma^2\langle (\log{w})^2\rangle \ \ .
\end{equation}
Figure \ref{fig04} shows the pair $(c_N,\gamma)$ of the system
of interacting networks addressed in Fig.~\ref{fig03}. One clearly
sees a linear dependence with a negative slope, $(\frac{dS}{d\gamma})_N<0$, 
which can be understood since $\gamma$
plays the role of the inverse of a temperature.
\begin{figure}[t]
\centerline{\includegraphics[width=0.99\linewidth]{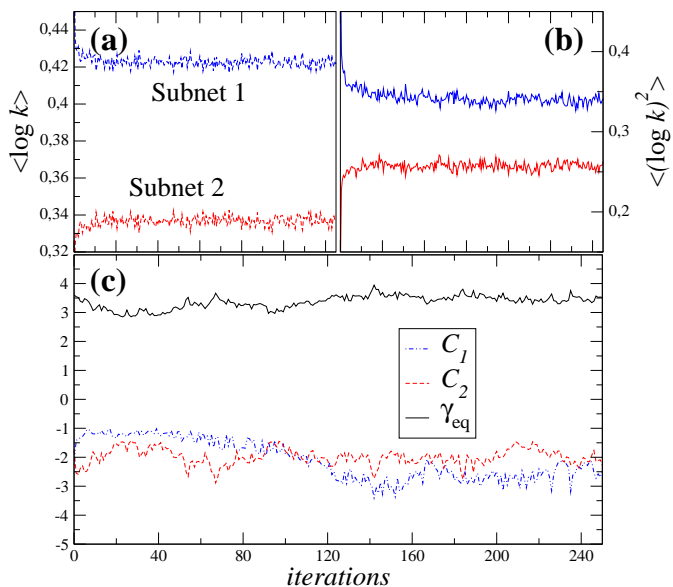}}
\vspace*{8pt}
\caption{\protect (Color online)
                  Computing equilibrium exponents and heat capacity of 
                  scale-free networks:
                  {\bf (a)} first and 
                  {\bf (b)} second moments of the energy 
                  $e_i\equiv \log{w_i}=\log{k_i}$ in two separated 
                  unweighted scale-free 
                  networks (blue and red).
                  {\bf (c)} The corresponding heat capacities, $c_1$ and 
                  $c_2$ (Eq.~(\ref{eq::heat.capacity})) and equilibrium 
                  exponent 
                  $\gamma_{eq}$ (Eq.~(\ref{eq::gammaeq})).} 
\label{fig05}
\end{figure}

Fifth, one can predict the value of the exponent $\gamma_\mathrm{eq}$
when two networks of different degree exponents, $\gamma_1$ and
$\gamma_2$, interact. Due to the interaction, internal links of each
network are rewired to connect to the other network. The variation of
energy $E_l$, which accounts for the overall variation of logarithmic
of the number of connections in network $l={1,2}$ obeys 
the relation,
\begin{equation}\label{eq::flow}
\Delta E_l= c_l\left(\gamma_\mathrm{eq}-\gamma_l\right) \ \ ,
\end{equation}
where $c^{(l)}$ is the heat capacity of network $l$. 
As the number
of links is conserved, $\Delta E_1=-\Delta E_2$. Consequently,
\begin{equation}\label{eq::gammaeq}
\gamma_\mathrm{eq}=\frac{c_1\gamma_1+c_2\gamma_2}{c_1+c_2} \ \ .
\end{equation}
Indeed, Eq.~(\ref{eq::gammaeq}), together with 
Eq.~(\ref{eq::heat.capacity}), provide a simple way for predicting
the equilibrium exponent of two scale-free networks that merge.

Figure \ref{fig05}a and \ref{fig05}b shows respectively the 
evolution of the quantities
$\langle \log{\gamma}\rangle$ and $\langle (\log{\gamma})^2\rangle$
in Eq.~(\ref{eq::heat.capacity}) for computing the heat capacity,
$c_1$ and $c_2$ in each network, which we plot in Fig.~\ref{fig05}c
together with $\gamma_{eq}$ calculated from Eq.~(\ref{eq::gammaeq}).
The result for $\gamma_{eq}$ are consistent with the ones plotted in
Fig.~\ref{fig03}.

\section{Discussion and conclusions}

In summary,  we presented an equilibrium 
description of scale-free networks with a growing number of
connections. A thermostatistical framework
was established, defining canonical and microcanonical ensembles of
power-law distributions.  With an algorithm for microcanonical 
ensembles, we showed that the merging of two scale-free networks, 
with different exponents, evolves towards an equilibrium power-law 
distribution. 
Moreover, we show how to predict the equilibrium exponent when 
two scale-free networks merge.

We have replaced, through rescaling, the growing system (growing 
$k$), which has an influx of energy (increasing $\epsilon_j$) by a 
corresponding ``isolated'' system which obeys thermostatistical laws. 
As should be noted, a similar choice for the energy has been
made in the  context of Bose-Einstein condensation in
networks\cite{Bianconi2001}.

Future work might focus on the geometrical properties of this new
ensemble of networks.
We consider a statistical ensemble of networks, with 
a time-dependent weight for each node, which is used to define 
an equivalent
to the energy of one microstate. We then
prescribe an iterative re-weighting
procedure, such that a functional, which contains
the analogue of the entropy, is maximized, while
constraining the average total energy and number of nodes.
We present expressions for the thermostatistical variables. 
From this, we propose the existence of
canonical and microcanonical ensembles, which are
scale-free networks with maximum entropy. 
We employ rescaling of the number of links, which is
not necessarily time invariant, and we focus on networks with a
constant number of nodes\cite{joaothesis}.

This new framework might provide insight into other problems. For
example, recently an interesting study of the Romenian wealth
distribution revealed that the evolution of the wealth distribution
yields a power law robust against external forcing and perturbations
(changes in policy, tax laws, ...)~\cite{zoltan}. The size of the
Romanian society is considered to be approximately constant in time, 
but the 
links (economical interactions) are dynamic.  In the spirit
of the framework we describe here, one might speculate that the
robustness of the power-law distribution is characteristic of an
``equilibrium'' state as the one reported here.
Well known examples of robust power-law distributions, with
approximately constant exponent in time, are also the Internet and the WWW.
In this case, the number of nodes is dynamic. A generalization of
our framework to networks with varying number of nodes might also help
understanding such robustness.
 


\acknowledgments

The authors thank Nuno Silvestre,
Nelson Bernardino and Jos\'e Maria Tavares
for useful discussions and
(PGL) German Environment Ministery under 
the project 41V6451, as well as
{\it Funda\c{c}\~ao para a Ci\^encia e a Tecnologia (FCT)} 
through PEst-OE/FIS/UI0618/2011 (FR and PGL),  
SFRH/BPD/65427/2009 (FR), 
and  {\it Ci\^encia 2007} (PGL), and from 
{\it FCT and Deutscher Akademischer Auslandsdienst (DAAD)} 
through {\it DRI/DAAD/1208/2013} (FR and PGL).


\bibliography{nbodymodelRefs}   

\end{document}